\documentclass[sigconf]{acmart}
\usepackage{graphicx}
\usepackage{subfigure}

\AtBeginDocument{%
  \providecommand\BibTeX{{%
    \normalfont B\kern-0.5em{\scshape i\kern-0.25em b}\kern-0.8em\TeX}}}

\copyrightyear{2024}
\acmYear{2024}
\setcopyright{rightsretained}
\acmConference[FDG 2024]{Proceedings of the 19th International Conference on the Foundations of Digital Games}{May 21--24, 2024}{Worcester, MA, USA}
\acmBooktitle{Proceedings of the 19th International Conference on the Foundations of Digital Games (FDG 2024), May 21--24, 2024, Worcester, MA, USA}
\acmDOI{10.1145/3649921.3649996}
\acmISBN{979-8-4007-0955-5/24/05}

\begin{document}

%%
%% The ''title'' command has an optional parameter,
%% allowing the author to define a ''short title'' to be used in page headers.
\title{Snake Story: Exploring Game Mechanics for Mixed-Initiative Co-creative Storytelling Games}

%%
%% The ''author'' command and its associated commands are used to define
%% ''authornote'' and ''authornotemark'' commands

\author{Daijin Yang}
\email{yang.dai@northeastern.edu}
\affiliation{%
  \institution{Northeastern University}
  \streetaddress{360 Huntington Ave.}
  \city{Boston}
  \state{MA}
  \country{USA}
  \postcode{02115}
}

\author{Erica Kleinman}
\email{e.kleinman@northeastern.edu}
\affiliation{%
  \institution{Northeastern University}
  \streetaddress{360 Huntington Ave.}
  \city{Boston}
  \state{MA}
  \country{USA}
  \postcode{02115}
}

\author{Giovanni Maria Troiano}
\email{g.troiano@northeastern.edu}
\affiliation{%
  \institution{Northeastern University}
  \streetaddress{360 Huntington Ave.}
  \city{Boston}
  \state{MA}
  \country{USA}
  \postcode{02115}
}

\author{Elina Tochilnikova}
\email{e.tochilnikova@northeastern.edu}
\affiliation{%
  \institution{Northeastern University}
  \streetaddress{360 Huntington Ave.}
  \city{Boston}
  \state{MA}
  \country{USA}
  \postcode{02115}
}

\author{Casper Harteveld}
\email{c.harteveld@northeastern.edu}
\affiliation{%
  \institution{Northeastern University}
  \streetaddress{360 Huntington Ave.}
  \city{Boston}
  \state{MA}
  \country{USA}
  \postcode{02115}
}

\renewcommand{\shortauthors}{Yang, et al.}

\begin{abstract}

%Play is intrinsically linked to creativity. With the advancement of generative AI, we now have mixed-initiative interactions that bolster this relationship. In these interactions, humans and AI collaboratively contribute, enriching both creative endeavors and playful experiences. Contrary to previous gamified mixed-initiative co-creation tools and mixed-initiative co-creation games, which typically position co-creativity as the central and sole experience in their design, there is an untapped potential for organically blending co-creation with other game mechanics to craft unique and immersive gameplay experiences.
Mixed-initiative co-creative storytelling games have existed for some time as a way to merge storytelling with play. 
However, modern mixed-initiative co-creative storytelling games predominantly prioritize story creation over gameplay mechanics, which might not resonate with all players. As such, there is untapped potential for creating mixed-initiative games with more complex mechanics in which players can engage with both co-creation and gameplay goals.
To explore the potential of more prominent gameplay in mixed-initiative co-creative storytelling games, we created \textit{Snake Story}, a variation of the classic \textit{Snake} game featuring a human-AI co-writing element. To explore how players interact with the mixed-initiative game, we conducted a qualitative playtest with 11 participants. Analysis of both think-aloud and interview data revealed that players' strategies and experiences were affected by their perception of \textit{Snake Story} as either a collaborative tool, a traditional game, or a combination of both. Based on these findings, we present design considerations for future development in mixed-initiative co-creative gaming.

\end{abstract}




%% A ''teaser'' image appears between the author and affiliation
  %\includegraphics[width=\textwidth]{Figures/Picture1.png}
  %\caption{\textit{Snake Story} Gameplay Flow}
  %\Description{}
  %\label{fig:teaser}


\maketitle

\section{Introduction}

Mixed-initiative co-creative storytelling games~\cite{mixedinitiativegame-wawlt} belong to a game genre that amplifies the playfulness of story creation through gameplay elements~\cite{casualcreators}. In these games, players interact with game systems and AI in a collaborative manner to generate a story through play~\cite{1001night, mixedinitiativegame-wawlt, mixedinitativegames-buddy, xi-etal-2021-kuileixi, mixedinitiativegame-looseend}. For example, in ''\textit{Why Are We Like This?}'' by Kreminski et al.~\cite{mixedinitiativegame-wawlt}, a rule-based AI lays the groundwork for a story world, complete with foundational characters, relationships, and backstories before players dive into the game. The AI then continues to present brief action ideas, which players can choose and expand upon through their own narratives. These player decisions dynamically influence the relationships between characters and the evolving histories of the story world.

In many ways, these games are a form of creative writing support, facilitating a mixed-initiative interaction~\cite{mixedinitiativeinterface} where humans and AI collaboratively craft stories. As such, they have the potential to democratize the creative writing experience for more non-serious creators~\cite{casualcreators}, who may place value on the playful and enjoyable nature of the process, which is then amplified by the game rather than the quality of the final written product. 

That being said, the existing state of the art for mixed-initiative co-creative storytelling games places the generation of the story as the core, and often sole, element of the gaming experience. For example, ``\textit{Why Are We Like This}'', introduced above, permits the player to select actions and dialogue for the characters almost like a director in a play~\cite{mixedinitiativegame-wawlt}. From a narrative game standpoint, this is an appropriate design approach. And from a broader perspective, it is likely the result of co-creative storytelling and gameplay mechanics conceptually being at odds with one another.
As one can imagine, having to react to time-sensitive ``twitch'' mechanics \cite{brathwaite2009challenges} makes it challenging for players to consider appropriate narrative progression or dialogue. However, simply accepting this as a fact disregards the potential for integrating co-creative processes within broader or more mechanically complex gaming experiences where co-creation is not the sole objective. In addition to creating novel and unique co-creative gaming experiences, such an approach may broaden the appeal of mixed-initiative co-creative storytelling games to a wider audience and, as explained above, make creativity more approachable.

To explore this possibility, we pose the following research questions:
1) How can we blend creative collaboration with more prominent game mechanics?
2) How do players engage with games that integrate both gameplay and co-creative elements?

To answer the first question, we adopted a research-through-design approach~\cite{rtd} and designed and developed \textit{Snake Story}, blending the traditional \textit{Snake} game with a GPT-3-based co-writing system. In the game, players use the snake's movement to select AI-generated text segments to add to an ever-growing narrative. As the game progresses, the player must balance the survival of the growing snake with the quality and coherence of the story, effectively combining the storytelling element with more complex gameplay mechanics and clear gameplay goals.

To answer the second question, we conducted a qualitative user study with 11 participants, collecting think-aloud, interview, and observation data. Analysis of the data revealed that players experienced limited authorship in co-creating stories within the game. Additionally, players demonstrated varied strategies and perceptions regarding their role within the gameplay experience, with there being a relatively clear breakdown of participants seeing themselves as either a (1) \textit{player}, (2) \textit{writer}, or (3) \textit{reader} as a consequence of experiencing the gameplay offered by our game. Based on these results, we discuss how to design future games like \textit{Snake Story} to better appeal to each role type. In all, we summarize our key contributions as follows:
\begin{itemize}
    \item [1)] We present a unique GPT3-based mixed-initiative co-creative storytelling game that mixes the goal of creating a story with more complex gameplay mechanics and explicit gameplay goals.
    \item [2)] Based on the results of a user study, we present insights into how players experienced the game and perceived their role in relation to the storytelling and gameplay goals and provide recommendations for the design of future games that seek to follow the same approach as \textit{Snake Story}.
\end{itemize}

\section{Related Work}

\subsection{Ruled-AI-Based Mixed-Initiative Co-Creative Storytelling Games}
The intrinsic playfulness of casual creativity~\cite{casualcreators, wawlt2} lies at the center of mixed-initiative co-creative storytelling games. This genre leverages human curiosity for co-creation, enhancing the writing experience with simple game mechanics. Predominantly, rule-based AI serves as the key technological underpinning in these games.

An example is IMPROV, designed by Kybartas et al., based on GLUNET, a semantic database containing associations between words~\cite{kybartas2015semantic}. 
In the game, players initially select a few words from the semantic database, and then a rule-based AI generates a story's opening and closing lines based on these players' selections. The player's goal is to complete the story through their own writing. Nevertheless, due to the lack of consistency checks and a world model, it's easy for players to create stories that lack meaningful coherence.

In Samuel et al.'s Writing Buddy~\cite{mixedinitativegames-buddy}, this issue is partially addressed by generating a story outline for players. The game is driven by more complicated rule-based AIs: the social simulation system Ensemble~\cite{samuel2015ensemble} and the play trace analyzer Playspecs~\cite{osborn2015playspecs}.
In this game, players strategically piece together a series of brief, pre-defined narrative goals to form the framework of a story. This narrative skeleton evolves as players make choices, ensuring that each decision shapes the unfolding tale. Following the construction of this outline, players can flesh out these narratives with rich, detailed prose.  

Another solution to this issue is to offer players a dynamically changing narrative world that provides specific action suggestions based on player interactions and continuously adapts the narrative in real time. For example, WAWLT by Kreminski et al.~\cite{kreminski2019felt}, as mentioned in the introduction section, employs Felt, a rule-based AI, to create initial settings and character dynamics. It offers action suggestions for player expansion, dynamically reshaping character relationships and histories. However, a playtest revealed that the action suggestions were occasionally inappropriate, leading to players losing direction in the game.
Building upon Writing Buddy and WAWLT, Kreminski et al.~\cite{mixedinitiativegame-looseend} also developed Loose Ends to counteract the aimlessness seen in the previous systems while still upholding the sense of co-authorship for players. 
Loose Ends differentiates itself by introducing more complex narrative goals that involve sequences of events with the possibility to include specific characters and constraints, enhancing personalization and depth in storytelling. Furthermore, unlike Writing Buddy, which merely directs the story toward player-set goals, Loose Ends' AI actively proposes new goals to maintain coherence and build upon the existing story, effectively acting as a collaborative writing partner.

In contrast to the above approach of positioning players as primary creators, Reed et al. developed Ice-Bound~\cite{reed2014ice}. In this game, players assume the role of an assistant to an AI writer, altering the overall narrative direction by selecting from various rule-based generative storytelling suggestions.

Beyond AI providing suggestions based on player actions, which players then select and write to alter a narrative, prior work has explored other forms of mixed-initiative mechanics. 
For instance, in TaleBox by Castano et al.~\cite{castano2016talebox}, an AI system powered by GLUNET~\cite{kybartas2015semantic} generates vocabulary cards that align with the game's evolving narrative objectives. Players strategically use these cards to craft actions that advance toward these set goals. The AI assesses the players' actions for coherence, allowing only those that adhere to a predefined standard. The game operates on a turn-based system, where players receive different vocabulary cards each turn. Cards not utilized by the end of a turn are discarded. Nevertheless, the lack of playtest reports makes it unclear what impact these mechanics have on the players' writing and gameplay experiences.
Building on the foundation of TaleBox, Perez et al. designed TaleMaker~\cite{talemaker}, a multiplayer, card-based storytelling game. Initially, one player sets the theme of the story. In each subsequent turn, an AI-based on WordNet~\cite{miller1995wordnet} provides players with eight vocabulary cards related to the current story, chosen from categories selected by the players (such as plants, food, tools, etc.). Players then use these vocabulary cards to form a sentence. Afterward, all players vote on the constructed sentences, and the one with the most votes is incorporated into the story. Quantitative analysis clearly shows that the stories generated are of high quality, and the game is generally well-received by players. However, despite acknowledging in the discussion section that players have mixed feelings regarding the game mechanics' influence on creative output, the authors did not pursue a comprehensive study to delve into the nuanced player responses to the prominent game mechanics.

\subsection{Language-Model-Based Mixed-Initiative Co-Creative Storytelling Games}

The advent of language models (LM) ushers in a new era for mixed-initiative co-creative storytelling games.
Previously, the tendency to use rule-based AI as the core for story generation stemmed from the inability of language models to generate coherent writing suggestions and narratives, which could potentially distract players~\cite{mixedinitiativegame-wawlt}. However, recent advancements in language models, such as GPT-2~\cite{gpt2} and GPT-3~\cite{gpt3}, have been demonstrated to produce text indistinguishable from human-written content~\cite{distinguishablegpt2, distinguishablegpt3}. This development has opened up new possibilities for applying LMs in mixed-initiative co-creative storytelling games.
LMs transcend the limitations of rule-based AIs that provide mere story outlines, instead delivering rich, prose-level narrative contributions. 
This capability opens up a boundless landscape of narrative possibilities, granting players the freedom to navigate and shape an infinite storytelling universe.

For example, AI Dungeon~\cite{AIDungeon}, a mixed-initiative co-creative storytelling game developed by Latitude, leverages LM to create an infinitely branching narrative, allowing players to embark on unique adventures limited only by their imagination. In this text-based game, players can type in any action or dialogue they can think of, and the AI dynamically generates a response, crafting a story in real time that adapts to the player's inputs. 
The game blurs the line between player and creator, fostering an immersive exploratory experience driven by players' curiosity regarding the unfolding narrative. 

However, as a game, AI Dungeon falls short in offering incentives to players, relying solely on their initiative for exploration. This approach can lead to a swift decline in players' engagement and enthusiasm~\cite{xi-etal-2021-kuileixi}.
To enhance LM-based mixed-initiative co-creative storytelling games beyond just exploration, simple game mechanics are incorporated to maintain player engagement and promote a state of flow. 
Xi et al.~\cite{xi-etal-2021-kuileixi} introduced KuiLeiXi, an AI-powered chatting game embedded within the romance simulation mobile game Yu Jian Love~\cite{Yujianlove}. In the game, players interact with a GPT-2-based AI.
Through conversation, players steer the AI towards producing text that achieves pre-defined game goals, such as consenting to reduce the price of an item. 

Sun et al.~\cite{sun2022bringing, 1001night} introduce more mechanics into LM-based mixed-initiative co-creative storytelling games. In their game 1001 Nights, players navigate storytelling and battle phases. Playing as Shahrzad, a character in the traditional ``1001 Nights'' story, they guide an AI King in creating tales tied to four distinct weapons. After making the stories, they fight the King with these weapons. The ultimate goal is to reforge Shahrzad's destiny, diverging from her classic narrative.

Nevertheless, current mixed-initiative co-creative storytelling games, including 1001 Nights, primarily emphasize co-creativity, with the co-writing aspect often being the sole focus. In 1001 Nights, for instance, players are assured victory in the battle phase, a feature that simplifies the gaming experience. Such an approach does not fully explore the integration of co-creative storytelling into more diverse gaming elements, where narrative generation is not the only goal. There remains an opportunity for further research on how to incorporate more intricate gameplay mechanics into mixed-initiative co-creative storytelling games.

\section{Snake Story}

\begin{figure*}[htbp]
\centering
\subfigure[In-game scene 1. (A) ``\textit{Snake}'' game part; (B) ``Story'' part; (C) timer; (D) Generated Texts with corresponding candy indicator; (E) life point indicator. Players can choose to eat one candy, and the corresponding text segment will be added to the story.]{
\includegraphics[width=3.2in]{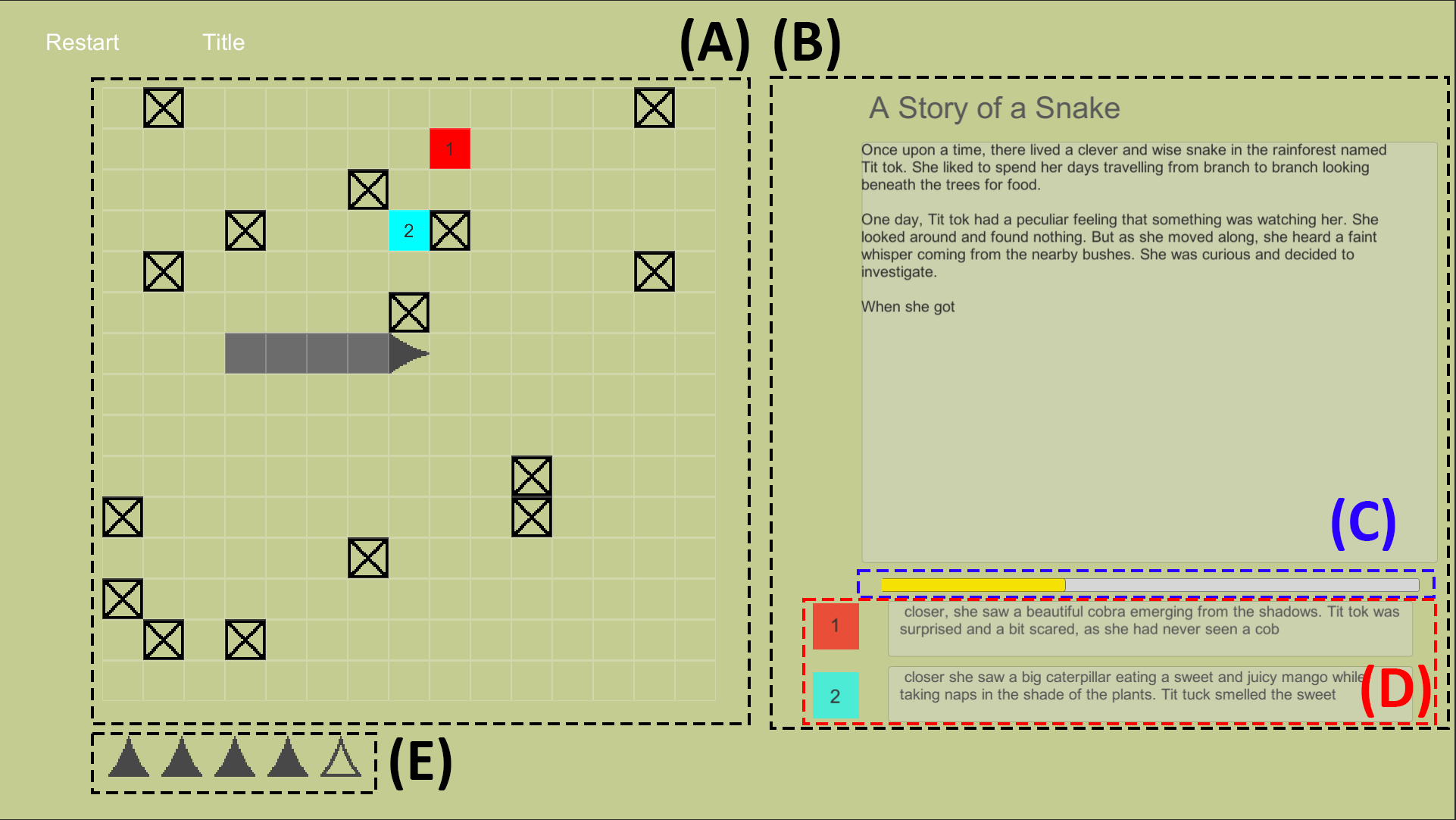}
\label{fig:ingameinterface}
}
\quad
\subfigure[In-game scene 2. If the players pick up a blue candy in the last round, a special yellow candy will appear on the map that allows players to input their own text in (F).]{
\includegraphics[width=3.2in]{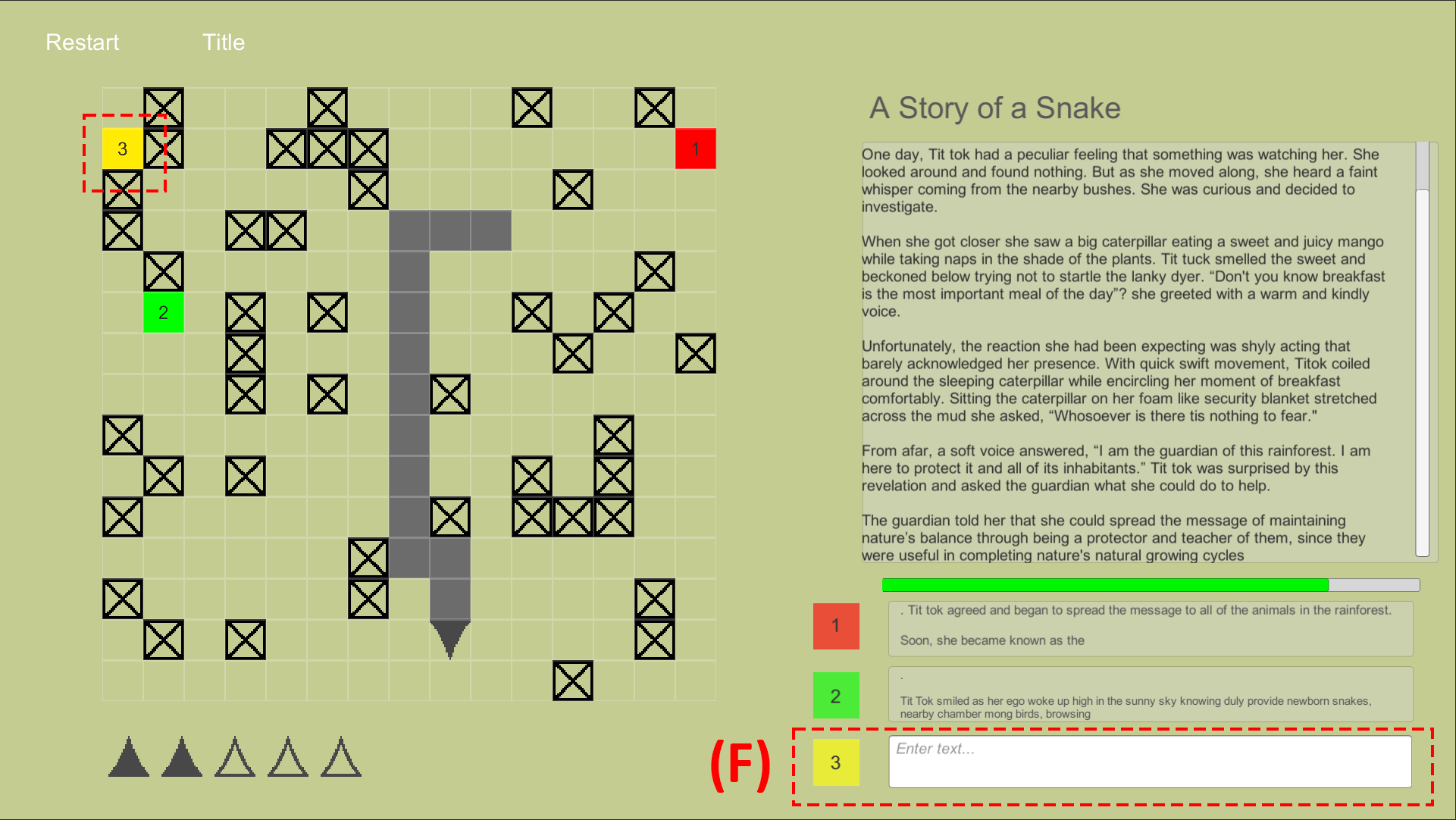}
\label{fig:ingameinterface2}
}
\subfigure[End-game scene. The game will end if players run out of life points, and this end scene will appear.]{
\includegraphics[width=3.2in]{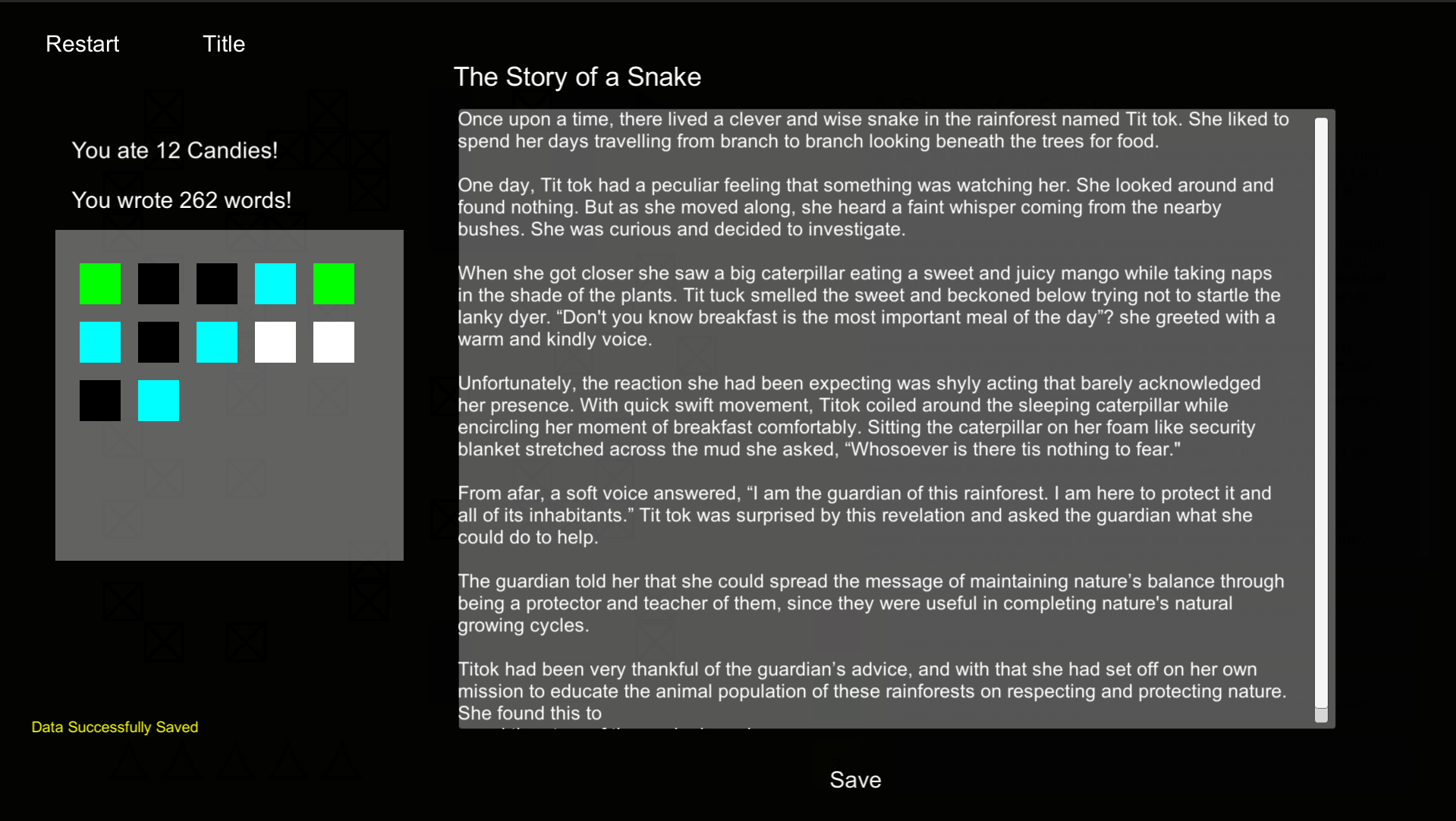}
\label{fig:endgameinterface}
}
\caption{\textit{Snake Story} Scenes.}
\end{figure*}

To study how mixed-initiative co-creation can work with more complex mechanics, we created \textit{Snake Story}, which combines a GPT3-based mixed-initiative co-writing system with the \textit{Snake} game. We adopted a research-through-design approach~\cite{rtd}, focusing on generating answers through our design efforts rather than examining solutions proposed by others. This approach was particularly necessitated by our interest in the use of GPT in games, an area that has not yet been extensively investigated~\cite{llm4game}.

The \textit{Snake} game is a classic video game concept where the player controls a line (representing a snake) that grows in length~\cite{SnakeGame}. The primary objective is to navigate the line around the play area to pick up food (causing the snake to grow) while avoiding collisions with the walls and the snake's own ever-expanding body. Each time the snake eats a piece of food, typically depicted as a dot or apple, it grows longer, making the game increasingly challenging as time goes on. The game continues until the snake collides with itself or the wall.
We were inspired to use this game for our design because: 
\begin{itemize}
    \item [1)] Gameplay that demands rapid decision-making (sometimes referred to as ``twitch'' gameplay \cite{brathwaite2009challenges}) is not only missing from current mixed-initiative storytelling games but is seemingly at the opposite end of the spectrum, as we discussed in the introduction. Hence, we started our investigation with the game Snake, which is a classic example of this style of gameplay, to observe the outcomes of introducing such contrasting gameplay elements.
    \item [2)] The “Snake” game is a relatively simple, classic, and well-known game, so it does not take participants or new players a long time to learn the gameplay, and many may already know it. Thus, it appropriately represents rapid gameplay without introducing unnecessary complexity or a need for lengthy tutorials. 
    \item [3)] The snake growing over time can be seen as a metaphor for the story developing over time. As such, the non-narrative mechanics of the game mirrored the storytelling element.
\end{itemize}

As shown in Fig.~\ref{fig:ingameinterface}, \textit{Snake Story} consists of two main parts: the \textit{Snake} game part (A) and the \textit{Story} mixed-initiative co-writing part (B). In the \textit{Snake} game part, players will navigate the snake (grey squares as its body with a triangle as the head) on the map (white grids) to collect candies (single-colored squares with a number) and avoid obstacles (black squares with a cross) as well as its own body. Collecting the candies is how the player adds text to the story, which we describe further below. The snake has a maximum of five life points, represented by the number of grey triangles at the bottom left of the interface (E). As with the original Snake game, the snake will lose one life point when its head hits an obstacle or its own body. When the snake runs out of life, the game will end.

\begin{figure*}[!htbp]
\centering 
\includegraphics[width=1\textwidth]{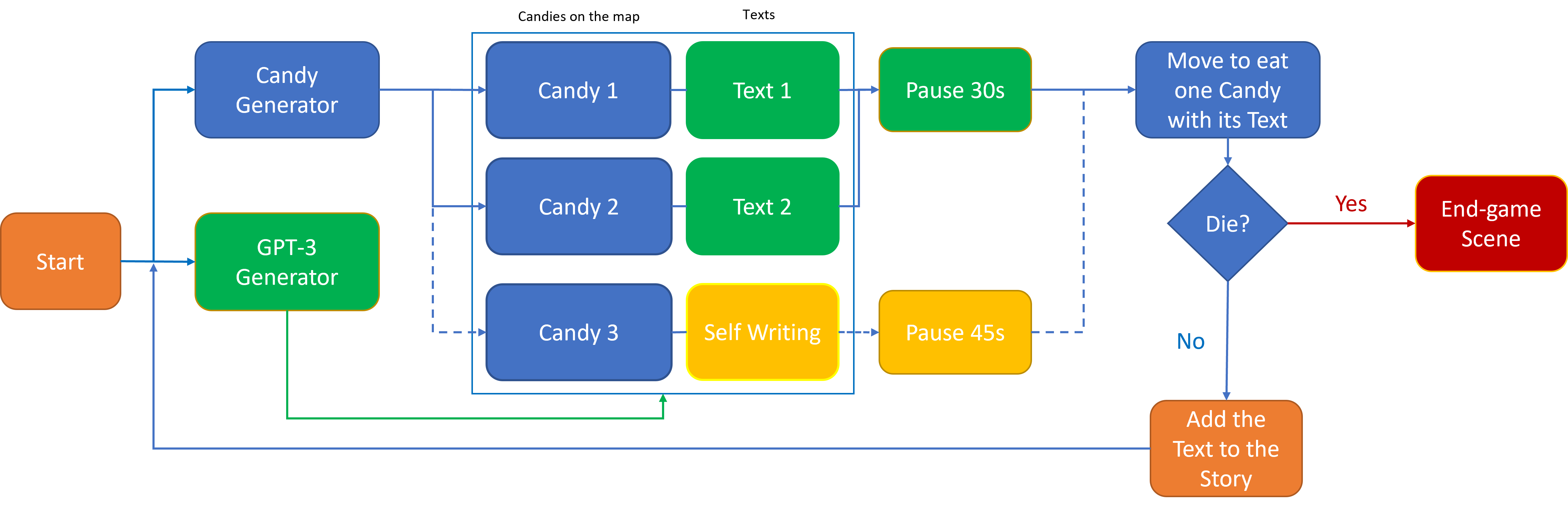} 
\caption{\textit{Snake Story} Gameplay Flow. \textit{Snake Story} is round-based. At the start of each round, candies labeled with the numbers 1 and 2, and occasionally 3 under special conditions, will appear on the map. Each candy is associated with a unique text segment generated by GPT-3. The game will then momentarily pause, allowing players time to read these segments. Subsequently, players will move the snake to hit the chosen candy, thereby incorporating its corresponding text segment into their unfolding story.}
\label{fig:teaser}
\end{figure*}

\textit{Snake Story} is round-based. Each round ends with players eating one of the candies by navigating the snake's head to hit it. As shown in the flow chart Fig.~\ref{fig:teaser}, at the beginning of each round, two candies with numbers 1 and 2 will be generated at two random locations that are not occupied by obstacles or the snake.
In addition to adding one grey square to the snake to make it longer, the candies will have different special effects defined by their number. 
As shown in Table~\ref{T:candy}, candies with the number 1 will grant an effect from the first three neutral and negative effects (Red, Black, White), whereas candies with the number 2 will grant an effect from the second three positive and neutral effects (White, Blue, Green). The candy's color indicates its effect, which is randomly determined at the moment of generation. Yellow candies, associated with the number 3, can only be generated by picking up a blue candy in the previous round. In addition to the effects displayed in Table~\ref{T:candy}, picking up candies is how players add text to the story, which we will describe in detail next.

In the ``Story'' mixed-initiative co-writing part of the game, the title \textit{A Story of A Snake} on the top right side of the screen indicates the topic of the story. Below the title, the story-in-progress is displayed, which evolves as the players add story content while playing.
Each candy on the board is associated with a text segment generated by the \textit{text-davinci-003} GPT-3 model. When players select the corresponding candy by navigating the snake to it, they add that text to the story. The mapping of text to candy is visible to the players and can be seen on the bottom right of the screen (D).

Among all GPT-3 models, the \textit{text-davinci-003 model} was the state-of-the-art model in text completion at the time of our study in October 2022~\cite{GPTDocument}. The first two text segments in the game (i.e., the first two candies the player has to choose between as the play) are generated by the prompt ``\textit{Write a story of a snake:}'', and the text segments that come after are generated based on the story so far.  
The style of the text is determined by the \textit{temperature}~\cite{GPTDocument}, a parameter of GPT-3 that can control the output quality of the model. Here, we manipulate this parameter to generate different kinds of text. 
For text segments that will be assigned to candies with the number 1, the temperature is set to 0.6, making the output more focused and coherent but potentially less creative and more repetitive. For example: ``\textit{The snake stayed there for many years while the house was slowly remodeled and restored to its former glory. As time passed by,}'' or ``\textit{Once upon a time, there lived a snake in the depths of the forest. She was an old and wise creature who rarely ventured out from her}'' (both number 2 red candies). The style of the text, as generated by the model, does not interact with the coloring of the candy.
For text segments that will be assigned to candies with the number 2 (Blue, Green, or White), the temperature is set to 1.4, which will make the output more creative and random but potentially more nonsensical or off-topic. For example: ``\textit{the trees and hearing songs brought by the breeze. Occasionally, he even heard rumblings from land dreaders - potential adversaries ready to}'' (number 1 white candy) or ``\textit{Harry the snake loved to explore. Whether it was a backyard, intense desert landscape, or lengthy forest he enjoyed piercing through. One day though Harry}'' (number 1 green candy). Furthermore, players are not informed about the temperature difference between the two text selections.

The story is generated in real-time, with GPT taking the entire story as its input to generate new text segments for the players. The player's input is directly added to the story, meaning it will also be considered part of the input for GPT to generate new content in the next round. The generation process takes about 2-5 seconds.

Due to the random generation of the text segments, if players want to select certain segments that better fit their story or storytelling goals, they will sometimes have to select candies with negative effects, potentially making the game more difficult, as seen in Table~\ref{T:candy}. For example, a really good segment may be attached to a black candy, which will add obstacles when selected but ultimately provide a more desirable continuation to the story. Alternatively, the player may need to select a green candy to guarantee their continued success in gameplay, but this may add an undesirable addition to the story. In essence, this turns the game into a resource management challenge, where players must balance the nature of the story with their ongoing success in gameplay.

To make the gameplay flow fluently, we limit the inputs from players, and the primary interaction is to read and select from the two AI-generated text segments to add to the story. At the beginning of each round, when the candy is generated, and the text segments appear on the screen, the game will pause for 25 seconds for players to read the generated texts. Under the square that displays the current story, there is a slide bar that serves as a timer (C), which will begin counting down after the generated text appears. Players will then proceed with gameplay as described above. Players can only give their own input if they pick up the special yellow candy. If the player picks up a yellow candy, the game will pause for 45 seconds to allow players to input their own text. The text box that facilitates this appears in the bottom right corner (F), as seen in Fig.~\ref{fig:ingameinterface2}.

If the player runs out of life points for the snake, the game ends, and GPT-3 will give the story an ending within 80 words by connecting the story to the prompt ``\textit{, and the story of the snake ends}''. As shown in Fig.~\ref{fig:endgameinterface}, an ending scene will appear to display the final story, the word count, and the candies eaten. This is the only way for the game to end.

\begin{table}[!htbp]

\caption{Candy Mechanics Details. The generated candies on the map will have one of the mechanics in the table based on their number. They will also be associated with one text segment. Number 1 candies (Red, Black, White) will be associated with a more coherent text segment; number 2 candies (White, Blue, Green) will be associated with a more creative text segment; and the number 3 (Yellow) candy will allow players to input their own texts.}
\begin{tabular}{|c|c|c|p{3cm}|}
\hline
\textbf{Color} & \textbf{Number} & \textbf{Effect Type} & \multicolumn{1}{c|}{\textbf{Effect}}  \\ \hline \hline
Red & 1 & Negative & Reduce 1 health. \\\hline
Black  & 1 & Negative & Add 3 additional obstacle blocks on the map.\\\hline
White & 1 & Neutral & No effect.\\ \hline\hline
White & 2 & Neutral & No effect.\\\hline
Blue & 2 & Positive & Generate a yellow candy next round\\\hline
Green& 2 & Positive & Restore 1 health.\\ \hline \hline
Yellow & 3 & Special & Pause the game 20 seconds longer to allow players to write their own text for this candy. \\ \hline
\end{tabular}
\label{T:candy}
%}
\end{table}

\section{User Study}

To research how different participants interact with \textit{Snake Story}, 11 game design students (referred to as P1-P11) from a university in the United States were recruited through social media. We aimed to recruit a sample with diverse experience with writing and AI, and this information for our participants is presented in Table~\ref{table:detail}.

In each experimental session, the participant was given a brief tutorial on the game mechanics and interface of \textit{Snake Story}. The participant was then asked to play the game until the snake died, which resulted in about 10-15 minutes of gameplay per participant. The participant was also required to engage in think-aloud protocols during gameplay, where they were asked to explain their decisions in terms of which text segments they selected and why. After gameplay, the participants participated in a semi-structured interview, lasting approximately 15-20 minutes, as they were asked to evaluate the story they wrote, explain their interaction strategies, and share their interaction experiences and role perceptions on AI and themselves. The interview questions are listed in Table~\ref{T:InterviewQ}.

\begin{table*}[!htbp]
\caption{Semi-structured interview questions.}
\begin{tabular}{|c|l|}
\hline
& \textbf{Text Evaluation}  \\ \hline \hline
1.& Could you please rate the overall quality of the story?\\\hline
2.& What do you think about the language/logic of the story?\\\hline
3.& How much control do you feel you have over this story?\\ \hline\hline
& \textbf{Interaction Strategies}\\\hline
4.& How do you decide your selections on texts and candies in \textit{Snake Story}?\\\hline
5.& What is your primary goal in \textit{Snake Story}?\\ \hline \hline
& \textbf{Experiences and Role Perceptions}\\ \hline
 6.&How do you feel about the game mechanics and the story creation in \textit{Snake Story}?\\\hline
 7.&How would you describe the role of yourself in \textit{Snake Story}?\\\hline
 8.&How would you describe the role of AI in \textit{Snake Story}?\\\hline
\end{tabular}
\label{T:InterviewQ}
%}
\end{table*}

The game captured and stored the text segments generated by GPT-3, as well as the participants' individual selections and the resulting stories. The screen and the audio were recorded during the experiment. The think-aloud and interview data were analyzed qualitatively following an open coding method~\cite{opencoding}. One researcher familiarized themselves with the data, coded the data, and then later grouped the codes into clusters and derived common themes emerging from the data. We discuss the results of this process in the next section. University IRB approved the protocol.

\section{Results}

\begin{figure}[htpb]

\begin{minipage}[b]{1.0\linewidth}
  \centering
  \centerline{\includegraphics[width=9.5cm]{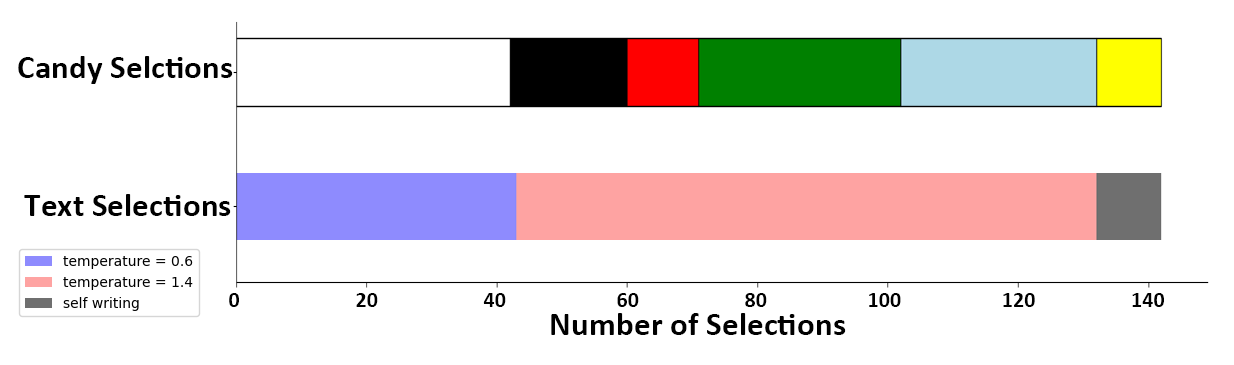}}

\caption{Statistics on Players' Text and Candy Selections. In the candy selection bar, colors correspond to the types of candies detailed in Table~\ref{T:candy}. The purple segment indicates the count of selected texts linked to the number 2 candies within the text selection bar. The pink segment denotes the count of selected texts associated with number 1 candies. Lastly, the gray segment shows how often players wrote their own text.}
\label{statistic}
\end{minipage}
\end{figure}

\subsection{Game Log Results}
As shown in Fig.~\ref{statistic}, the 11 participants played a total of 142 rounds (\textit{M} = 12.91, \textit{SD} = 4.50) in \textit{Snake Story}. In the \textit{Story} part, the text segments that were generated with a lower temperature (number 2 candies) (0.6) were selected 43 times (\textit{M} = 3.91, \textit{SD} = 1.98), while the higher temperature (number 1 candies) (1.4) texts were selected 89 times (\textit{M} = 8.09, \textit{SD} = 4.72). Participants chose to write their own words (yellow candies) 10 times (\textit{M} = 0.91, \textit{SD} = 1.83). The 11 stories had 272.64 words on average (\textit{SD} = 64.22).

In the \textit{Snake} game part, 91 white candies were generated, 42 of which were selected (46.15\%); 50 black candies were generated, 18 of which were selected (36.00\%); 47 red candies were generated, 11 of which were selected (23.40\%); 46 green candies were generated, 31 of which were selected (67.39\%); 47 blue candies were generated, 30 of which were selected (63.83\%); 40 yellow candies were generated, 10 of which were selected (25.00\%).
Three participants chose the yellow candy (associated with the human-authored story) in the game. P3 took one without contributing any content. P6 selected six, and P11 chose three yellow candies.
 
\subsection{Qualitative Results}
\subsubsection{Self Evaluation of the Co-writing}

In the \textit{Snake Story} study, participants generally viewed the co-authored narratives as linguistically proficient but lacking in logical consistency and coherence. While the language used by the AI was appreciated for its grammatical correctness and choice of words, participants noted a disconnect in the narrative flow. Participant 6 articulated this by stating, ``\textit{I felt the words the AI used were really nice, and there were hardly any grammar mistakes. But because I picked a lot of good candies (with less coherent texts), I don't feel like the story flows that well. }'' Furthermore, all participants perceived their authorial contribution to the co-written stories as neutral or minimal due to the mechanics in the ``\textit{Snake}'' game part. For instance, P7 explained ``\textit{Since the game's mechanics kind of keeps me from the text I want, I feel like I don't really have much control over the story.}''

\subsubsection{Play and Creation Strategies}

The mechanics in the \textit{Snake} game part influenced how participants decided to treat the generated text segments. Two out of eleven participants reported that they never read the generated text. For instance, P3 mentioned that ``\textit{Given that it's a game, I'm not particularly concerned about what the AI writes. My main focus is on the gameplay itself.}'' 
Five participants opted to give the generated texts a quick scan due to the limited pause time. To illustrate, P8 mentioned ``\textit{So basically, I just skim through the text real quick cause I also need to focus on figuring out how to get my snake to chow down on what I picked out for it at the same time.}''
On the other hand, some participants were more interested in reading the texts more thoroughly, and their gameplay strategies adapted to accommodate this. For example, during the 5th round, P11 commented, ``\textit{I think I can find a safe path for my snake to stay in, and then I can have extra time for reading the texts. Oh, this works!}''. 

Additionally, players' storytelling choices were influenced by their game state. In the rare case where two neutral white candies with no special effects appeared together (see Table~\ref{T:candy}), all participants were observed to choose the text they preferred. For instance, P11 said ``\textit{The second option does not make so much sense to me. I am going to choose the first.}''
The more common case was that two different candies were generated, and the preferred text piece was assigned to the candy with a worse effect. If the participants felt they were in a good game state (i.e., had a lot of health and not too many obstacles to navigate), they tended to ignore the candy effects and choose the text they preferred. For instance, in round 2 of the game, P8 stated ``\textit{That's (option 2) sounds too conceptual. I'm going for one even though it will reduce my health.}'' 
A rarer case was when the preferred text piece appeared attached to the candy, which had a positive effect; in this case, participants picked that candy without hesitation. P10 explained ``\textit{I like this Harry (the snake). It can also restore my health. I will go with the Harry story.}'' 

However, in most cases, participants would sacrifice their stories for easier gameplay. The green and blue candies (candies with positive effects) were particularly attractive to participants. To illustrate, P2 mentioned ``\textit{I would more likely go for the green candy to regain my lost HP and keep myself alive in the game for a bit longer.}''. By contrast, participants rarely chose black and red candies (candies with negative effects). For example, P7 mentioned that ``\textit{Even though the texts in the black candy are better, I'm not really keen on making the game more challenging. Plus, the white candy's text is good enough for me.}'' The priority to game mechanics became even stronger when the participant was about to die. For instance, P11 explained ``\textit{I really want to choose (option) 1, but I die if I choose 1. This is a very hard dilemma. I wish I could choose 1.}'' 

\begin{table*}[htpb]
\begin{minipage}[b]{1.0\linewidth}
\centering
\caption{The Individual Reports of 11 Players. \textbf{AI experiences}: N (No), Y (Yes); \textbf{Writing experiences}: R (rich), P (Poor), N (NO); Reading Strategies: S (Skim), T (Read Thoroughly), N (Never Read); \textbf{Play Strategies}: I (Ignore Writing), T (Make trade-offs); \textbf{Role Perception}: W (Writer), P (Player), R (Reader).}
\begin{tabular}{||c | c c c c c c c c c c c||} 
 \hline
 \textbf{Player Codes} & \textbf{P1} & \textbf{P2} & \textbf{P3} & \textbf{P4} & \textbf{P5}& \textbf{P6}& \textbf{P7}& \textbf{P8}& \textbf{P9}& \textbf{P10}& \textbf{P11}\\ [0.5ex] 
 \hline
\textbf{AI experiences} & N & Y & Y & N & N & Y & N & Y & N & Y & N\\ 
\textbf{Writing experiences} & R & P & N & R & R & P & N & R & N & N & N \\ 
\hline
\textbf{Reading Strategies} & S & T & N & N & S & S & S & S & T & T & T \\ 
\textbf{Playing Strategies} & T & T & I & I & T & T & T & T & T & T & T \\ 
\hline
\textbf{Role Perception} & W & R & P & P & R & W & R & W & R & R & R \\ \hline
  
\end{tabular}

\label{table:detail}
\end{minipage}

\end{table*}

\subsubsection{Role Perception and Player Experiences}
\label{result:roleandexperiences}

Analysis of our results revealed that the players perceived themselves to align with one of three different roles, which influenced how they perceived the game and felt about the experience.
\vspace{0.2cm}

\textbf{Role 1: Writers}\\
Three (P1, P6, P8) of the 11 participants emphasized, through their responses, that they were focused on the generation of the story. Based on this, we categorize these participants as ``writers'' who saw the game as a tool for writing a story. All of them found the gameplay unfavorable and considered the mechanics bothersome as they consistently interfered with their text options. For example, the time limit made the writing stressful, as illustrated by P8, who said ``\textit{It is not that fun to create stories with this (system). I do not have enough time to read and make decisions. That's kind of stressful.}''. Additionally, the candy mechanics confused these players. For example, P6 mentioned ``\textit{I feel like there are too many things I need to pay attention to in the game, and the texts I want to choose often appear on candies with negative mechanisms. It is very hard for me to create my own story.}''. Moreover, the death of the snake would often frustrate them, as illustrated by P6, who mentioned ``\textit{I'm not a fan of this game because I'm looking for a longer and more engaging story. Right now, it just seems way too easy to get killed in it.}''
\vspace{0.2cm}

\textbf{Role 2: Players}\\
Two participants (P3, P4) emphasized a focus on playing the game and surviving for as long as possible, ignoring the storytelling aspect. As such, we categorize these participants as ``players''. These participants liked the gameplay but felt the collaborative aspect superfluous. P4 explained this best by saying ``\textit{My attention gets captured by the game mechanics, and I’m more inclined to play the game rather than seriously consider the logic of the story. I’m interested in testing the rules of the game with the aim of winning, not crafting an interesting story. After all, playing a game and creating are two different things.}''
\vspace{0.2cm}

\textbf{Role 3: Readers}\\
The remaining six participants perceived ``\textit{Snake Story}'' as a blended experience with both gameplay and co-creation, where the gameplay was how they could continue to experience the story. We categorize these participants as ``readers'' due to their emphasis on being someone who experiences the story through continued gameplay but with a notable lack of emphasis on being the \textit{creator} of the story or responsible for guiding. In other words, they were ``reading'' the story, and their desire to continue reading became their motivation to play. To illustrate this, P2 said ``\textit{I keep picking the good candies to extend the snake's life since I'm eager to read more of the story. ... Yeah, I really think looking forward to the next part of the story could be one of the reasons people enjoy playing this game.}''

These participants also seemed to enjoy the challenges in the \textit{Snake} game part. The challenges served as motivators for participants to interact with the system and read the generated texts. As an example, P9 illustrated ``\textit{The game itself gives me a sense of urgency, and I find that this urgency is what makes it fun. It keeps me more engaged with the stories generated by the AI. Without that feeling of urgency, I would spend too much time trying to come up with stories, which would actually feel like a burden to me and take away from the enjoyment of reading.}''

\section{Discussion}
In this section, we discuss how the three roles our participants fell within (i.e., writers, players, and readers) may have been impacted by the game design. Furthermore, based on our results, we can see that each of the three roles approached the game with different objectives. Thus, different considerations need to be made to appeal to each role. As such, we discuss how we could possibly design games like \textit{Snake Story} from the perspective of each role.

\subsection {Designing for Writers}
The first role we saw, \textit{writers}, encompassed players with strong writing objectives and expected the gameplay to align with and support their writing goals. Writers had high demands on content quality and often expressed frustration when playing the game. This frustration was born from the fact that the gameplay goals and mechanics were often at odds with their creative aspirations and writing objectives. More specifically, writers exhibited a preference for higher-quality text segments, but the game did not always provide such segments or did not provide them in such a way that the player could afford to select them. For example, the player may have needed to select a lower quality segment to continue to live if they were on their last life. This conflict between gameplay goals and writing goals served as the main source of the \textit{writers'} frustration.

The writers' experience highlights an inherent conflict between creation and gameplay tension. Video games are designed to maintain player engagement by offering progressively challenging tasks that match the player's growing skill set~\cite{flowtheory, baumann2016flow}, creating a dynamic of tension and release integral to game design~\cite{schell2008artgamedesign}. However, this kind of tension does not align well with creative endeavors. In the realm of creation, tension tends to shift the emphasis towards quantity rather than quality~\cite{uncertainwriting}, which conflicts with the primary objectives of writers focused on crafting high-quality content. 

Based on these conclusions, if we are to design games like \textit{Snake Story} for writers, the gameplay systems should facilitate writing objectives rather than being at odds with them. For instance, in \textit{Snake Story}, we could assign better text segments to the candy with positive effects. The design could further emphasize the mixed-initiative co-writing system, offering appropriate AI interactions for \textit{writers}.
This includes editing and enhancing texts composed by humans~\cite{cowriteedit1, cowriteedit2}, filling in the narrative gaps within the skeletal frameworks of stories~\cite{cowritingfillgap, cowritingfillgap2}, generating ideas for narrative development~\cite{cowritingideageneration1, cowritingideagenerating3}, or guiding the direction of plots by generating coherent subsequent sentences or paragraphs~\cite{cowriting-aiasactivewriter, cowritingideageneration2}.
Ideally, the game system should act as a gamified extension of the writing process, functioning as an external motivator for writers to engage with the system~\cite{wiethof2021designinggamifiedmixed}. Game objectives should be harmonized with writing goals to avoid conflicts, and gameplay tensions should be minimized to enhance the writing experience.

\subsection{Designing for Players}
The second role we recognized, \textit{players}, encompassed participants who had no writing goals and were engaged with only the \textit{Snake} gameplay in \textit{Snake Story}. On the one hand, these players were easily engaged with the gameplay part of the game and, overall, enjoyed their experience playing. On the other hand, the absence of external writing goals means that these players were largely unmotivated to engage with the story generation system during gameplay. 

Overall, players demonstrate a limited interest in narratives and collaborative creation and are probably the audience that games like \textit{Snake Story} are least suited for. While this may not be the target audience for games like this, we can still think about design adjustments that may better capture their interest. 
One option would be to enhance the mechanism connection between gameplay and narrative. For instance, incentivizing \textit{players} through rewards for their narrative creations might be effective. This could involve offering gameplay advantages, like additional lives, in recognition of narrative successes, whether it be for coherence, length, rhythm, the development of a narrative arc, or adherence to genre norms. Such a strategy motivates players to focus on and value the narrative as a means to advance their gameplay experience.
Another option would be to try to use the design strategies described above or in the next section to try to guide or shift players towards a different role by better tempering or setting up their expectations~\cite{carstensdottir2021naked} for the game and perhaps influencing them to take on goals better aligned with the reader or writer role. 
Additional options may include acknowledging that players simply are not interested in narrative creation and that the best way to engage them with the game is not to force them to engage with the narrative. In other words, it is essential to provide players with the option to completely separate gameplay from collaborative creation, allowing them to independently enjoy the game without worrying about generating a story.

\subsection {Designing for Readers}
The final role we observed, \textit{readers}, encompassed players who were engaged with the narrative creation portion of the game but not from the standpoint of having strong writing goals. Instead, they wished to create longer stories to continue reading what happened next. Readers were motivated to continue playing the game as doing so was what allowed them to continue enjoying the story of the snake. Unlike writers, readers did not feel explicit frustration as their creative objectives did not conflict with their gameplay objectives, as surviving longer in \textit{Snake Story} means the AI generates longer stories. This connection between gameplay and creation transformed the \textit{Snake} gameplay into a driving force for the players' reading and helped them engage with both the gameplay and story creation elements of the game at the same time.

While it is difficult to conclude from our results alone, it may further be that readers felt a sense of alleviated responsibility in creating the story, instead viewing the AI as the active writer~\cite{cowriting-aiasactivewriter} and they themselves as only consumers. This may have helped them enjoy the story being created without feeling the diminished sense of authorship that those participants aligned more with a ``writer's'' role experienced. From this, we may be able to infer that readers are the ideal audience for games like \textit{Snake Story}.

Although the readers in our study were relatively pleased with their overall experience, we can still theorize ways to improve it further. The key to effectively crafting mixed-initiative storytelling games tailored for readers is to reinforce the connection between gameplay and the generation of the narrative to foster cohesive interaction. We could do this in several ways. However, we discuss two in particular. 

First, the game mechanics should be linked to the AI-generated text to undertake narrative functions. This can help to alleviate the traditional disconnection between game rules and fiction~\cite{conflict}, providing more immersive experiences for readers. For instance, if the generated plots revolve around a protagonist acquiring new abilities, the player could also gain new gameplay abilities, reflecting the development of the narrative in the gameplay mechanics. Of course, such an arrangement would require implementing constraints on AI-generated content, as AI-generated texts cannot be anticipated without imposing limitations, and it would otherwise be difficult to ensure that the mechanics evolve in relation to the generated narrative. Furthermore, leveraging other generative AI tools can be instrumental in creating real-time narrative-linked game assets. A notable example is 1001 Nights~\cite{1001night}, where ControlNet~\cite{Zhang_2023_ICCV_Controlnet} and a pixelization tool~\cite{pixelization} are employed to craft pixelated backgrounds that are thematically consistent with the collaboratively crafted story in battle scenes. 

Another option is to dynamically adjust the pacing~\cite{pacing} of the gameplay to match the progression of the generated narrative. For example, a common narrative structure~\cite{narrativestructure} includes \textit{Exposition}, \textit{Inciting Incident}, \textit{Rising Action}, \textit{Climax}, \textit{Falling Action}, \textit{Resolution}, and \textit{Denouement}. Based on this structure, we can increase the difficulty of the gameplay mechanics during the narrative's \textit{Rising Action}, reach the maximum intensity during the \textit{Climax}, and then release the tension during the \textit{Falling Action}. In doing so, we may be able to reflect the events of the story in gameplay without necessarily having to change the game mechanics themselves.

\section{Limitations}

Our research has identified various preferences and strategies among different types of players in mixed-initiative co-creative storytelling games, and based on these findings, we have proposed several design suggestions. However, our work is not without limitations. First, we selected students majoring in game design as our subjects. While this ensured high-quality feedback from a smaller dataset, it did not adequately represent the experiences of a broader range of casual players and non-players. In the future, we hope to expand our participant pool to include these unrepresented groups. 
Second, We acknowledge that in choosing \textit{Snake} as the basis for our game, we have created a very specific type of gameplay experience, one that only offers a singular goal and a limited number of mechanics. We acknowledge that games of different genres and levels of complexity, when incorporating co-creative storytelling elements, would create very different experiences and that the results we saw, specifically the roles that emerged, are the result of the experience we created. In future work, we will experiment with our mixed-initiative mechanics but with different game genres (e.g., shooter, platformer) to further explore and extend the player experience with gamified mixed-initiative storytelling.
Third, although players did not report any layout and text readability issues due to the game's user interface (UI) design, we acknowledge that this could be a problem, for which future work may want to inquire further about what UI best supports the player experience while allowing for creative storytelling to emerge organically.
Finally, we acknowledge that our sample consisted of 11 people. While this is a suitable number for qualitative and exploratory work, similar to what is seen in existing literature~\cite{halabi2019assessing}, we acknowledge this is a small sample, which means that our identified roles will need to be scrutinized with a larger (and more diverse) sample size in future work for generalizing the results of our exploratory inquiry.

\section{Conclusion}

We presented \textit{Snake Story}, a mixed-initiative co-creative storytelling game that integrated prominent gameplay objectives. In \textit{Snake Story}, players play the classic \textit{Snake} game while collaborating with GPT-3 to co-create a story about a snake's adventures. This unique gameplay experience requires players to balance their focus between the game's traditional objectives and their creative narrative ambitions.
In our study with 11 players, the complex interplay between gameplay mechanics and storytelling is brought to the forefront through the diverse roles players experienced---\textit{writers}, \textit{players}, and \textit{readers}. \textit{Writers}, charged with crafting the narrative, frequently found it challenging to fully engage with the gameplay aspect, while \textit{players}, primarily focused on gameplay, occasionally overlooking the narrative aspects. On the other hand, \textit{readers} experienced a more balanced interaction, appreciating both the gameplay and the story. 

The insights from \textit{Snake Story} suggest a nuanced approach to future game design in this genre. For \textit{writers}, the game environment should prioritize narrative creativity, with gameplay elements that support rather than hinder the storytelling process. \textit{Players}, in contrast, would benefit from a focus on immersive and engaging gameplay mechanics, where the narrative can serve as a complementary feature. For \textit{readers}, an ideal design would seamlessly blend narrative and gameplay, enhancing their overall experience. This approach, tailored to the specific preferences of different player types, points towards a future where mixed-initiative co-creative storytelling games can more effectively cater to a diverse gaming audience.

\bibliographystyle{ACM-Reference-Format}
\bibliography{sample-base}


\begin{thebibliography}{49}

%%% the Web address.

\ifx \showCODEN    \undefined \def \showCODEN     #1{\unskip}     \fi
\ifx \showDOI      \undefined \def \showDOI       #1{#1}\fi
\ifx \showISBNx    \undefined \def \showISBNx     #1{\unskip}     \fi
\ifx \showISBNxiii \undefined \def \showISBNxiii  #1{\unskip}     \fi
\ifx \showISSN     \undefined \def \showISSN      #1{\unskip}     \fi
\ifx \showLCCN     \undefined \def \showLCCN      #1{\unskip}     \fi
\ifx \shownote     \undefined \def \shownote      #1{#1}          \fi
\ifx \showarticletitle \undefined \def \showarticletitle #1{#1}   \fi
\ifx \showURL      \undefined \def \showURL       {\relax}        \fi
% invisible to TeX
\providecommand\bibfield[2]{#2}
\providecommand\bibinfo[2]{#2}
\providecommand\natexlab[1]{#1}
\providecommand\showeprint[2][]{arXiv:#2}

\bibitem[Ammanabrolu et~al\mbox{.}(2020)]%
        {cowritingfillgap}
\bibfield{author}{\bibinfo{person}{Prithviraj Ammanabrolu}, \bibinfo{person}{Ethan Tien}, \bibinfo{person}{Wesley Cheung}, \bibinfo{person}{Zhaochen Luo}, \bibinfo{person}{William Ma}, \bibinfo{person}{Lara~J. Martin}, {and} \bibinfo{person}{Mark~O. Riedl}.} \bibinfo{year}{2020}\natexlab{}.
\newblock \showarticletitle{Story Realization: Expanding Plot Events into Sentences}.
\newblock \bibinfo{journal}{\emph{Proceedings of the {AAAI} Conference on Artificial Intelligence}} \bibinfo{volume}{34}, \bibinfo{number}{05} (\bibinfo{date}{apr} \bibinfo{year}{2020}), \bibinfo{pages}{7375--7382}.
\newblock
\urldef\tempurl%
\url{https://doi.org/10.1609/aaai.v34i05.6232}
\showDOI{\tempurl}


\bibitem[Baumann et~al\mbox{.}(2016)]%
        {baumann2016flow}
\bibfield{author}{\bibinfo{person}{Nicola Baumann}, \bibinfo{person}{Christoph L{\"u}rig}, {and} \bibinfo{person}{Stefan Engeser}.} \bibinfo{year}{2016}\natexlab{}.
\newblock \showarticletitle{Flow and enjoyment beyond skill-demand balance: The role of game pacing curves and personality}.
\newblock \bibinfo{journal}{\emph{Motivation and Emotion}}  \bibinfo{volume}{40} (\bibinfo{year}{2016}), \bibinfo{pages}{507--519}.
\newblock


\bibitem[Biskjaer et~al\mbox{.}(2019)]%
        {uncertainwriting}
\bibfield{author}{\bibinfo{person}{Michael~Mose Biskjaer}, \bibinfo{person}{Jonas Frich}, \bibinfo{person}{Lindsay MacDonald~Vermeulen}, \bibinfo{person}{Christian Remy}, {and} \bibinfo{person}{Peter Dalsgaard}.} \bibinfo{year}{2019}\natexlab{}.
\newblock \showarticletitle{How Time Constraints in a Creativity Support Tool Affect the Creative Writing Experience}. In \bibinfo{booktitle}{\emph{Proceedings of the 31st European Conference on Cognitive Ergonomics}} (BELFAST, United Kingdom) \emph{(\bibinfo{series}{ECCE '19})}. \bibinfo{publisher}{Association for Computing Machinery}, \bibinfo{address}{New York, NY, USA}, \bibinfo{pages}{100–107}.
\newblock
\showISBNx{9781450371667}
\urldef\tempurl%
\url{https://doi.org/10.1145/3335082.3335084}
\showDOI{\tempurl}


\bibitem[Brathwaite and Schreiber(2009)]%
        {brathwaite2009challenges}
\bibfield{author}{\bibinfo{person}{Brenda Brathwaite} {and} \bibinfo{person}{Ian Schreiber}.} \bibinfo{year}{2009}\natexlab{}.
\newblock \bibinfo{booktitle}{\emph{Challenges for game designers}}.
\newblock \bibinfo{publisher}{Course Technology/Cengage Learning Boston, Massachusetts}.
\newblock


\bibitem[Brown et~al\mbox{.}(2020)]%
        {gpt3}
\bibfield{author}{\bibinfo{person}{Tom Brown}, \bibinfo{person}{Benjamin Mann}, \bibinfo{person}{Nick Ryder}, \bibinfo{person}{Melanie Subbiah}, \bibinfo{person}{Jared~D Kaplan}, \bibinfo{person}{Prafulla Dhariwal}, \bibinfo{person}{Arvind Neelakantan}, \bibinfo{person}{Pranav Shyam}, \bibinfo{person}{Girish Sastry}, \bibinfo{person}{Amanda Askell}, \bibinfo{person}{Sandhini Agarwal}, \bibinfo{person}{Ariel Herbert-Voss}, \bibinfo{person}{Gretchen Krueger}, \bibinfo{person}{Tom Henighan}, \bibinfo{person}{Rewon Child}, \bibinfo{person}{Aditya Ramesh}, \bibinfo{person}{Daniel Ziegler}, \bibinfo{person}{Jeffrey Wu}, \bibinfo{person}{Clemens Winter}, \bibinfo{person}{Chris Hesse}, \bibinfo{person}{Mark Chen}, \bibinfo{person}{Eric Sigler}, \bibinfo{person}{Mateusz Litwin}, \bibinfo{person}{Scott Gray}, \bibinfo{person}{Benjamin Chess}, \bibinfo{person}{Jack Clark}, \bibinfo{person}{Christopher Berner}, \bibinfo{person}{Sam McCandlish}, \bibinfo{person}{Alec Radford}, \bibinfo{person}{Ilya Sutskever}, {and}
  \bibinfo{person}{Dario Amodei}.} \bibinfo{year}{2020}\natexlab{}.
\newblock \showarticletitle{Language Models are Few-Shot Learners}. In \bibinfo{booktitle}{\emph{Advances in Neural Information Processing Systems}}, \bibfield{editor}{\bibinfo{person}{H.~Larochelle}, \bibinfo{person}{M.~Ranzato}, \bibinfo{person}{R.~Hadsell}, \bibinfo{person}{M.F. Balcan}, {and} \bibinfo{person}{H.~Lin}} (Eds.), Vol.~\bibinfo{volume}{33}. \bibinfo{publisher}{Curran Associates, Inc.}, \bibinfo{pages}{1877--1901}.
\newblock
\urldef\tempurl%
\url{https://proceedings.neurips.cc/paper/2020/file/1457c0d6bfcb4967418bfb8ac142f64a-Paper.pdf}
\showURL{%
\tempurl}


\bibitem[Bueno~Perez and Bidarra(2022)]%
        {talemaker}
\bibfield{author}{\bibinfo{person}{Mijael~Ricardo Bueno~Perez} {and} \bibinfo{person}{Rafael Bidarra}.} \bibinfo{year}{2022}\natexlab{}.
\newblock \showarticletitle{Mixed-Initiative Story Co-Creation with TaleMaker}. In \bibinfo{booktitle}{\emph{Proceedings of the 17th International Conference on the Foundations of Digital Games}} (Athens, Greece) \emph{(\bibinfo{series}{FDG '22})}. \bibinfo{publisher}{Association for Computing Machinery}, \bibinfo{address}{New York, NY, USA}, Article \bibinfo{articleno}{33}, \bibinfo{numpages}{13}~pages.
\newblock
\showISBNx{9781450397957}
\urldef\tempurl%
\url{https://doi.org/10.1145/3555858.3555876}
\showDOI{\tempurl}


\bibitem[Carstensdottir et~al\mbox{.}(2021)]%
        {carstensdottir2021naked}
\bibfield{author}{\bibinfo{person}{Elin Carstensdottir}, \bibinfo{person}{Erica Kleinman}, \bibinfo{person}{Ryan Williams}, {and} \bibinfo{person}{Magy~Seif Seif El-Nasr}.} \bibinfo{year}{2021}\natexlab{}.
\newblock \showarticletitle{” Naked and on Fire”: Examining Player Agency Experiences in Narrative-Focused Gameplay}. In \bibinfo{booktitle}{\emph{Proceedings of the 2021 CHI conference on human factors in computing systems}}. \bibinfo{pages}{1--13}.
\newblock


\bibitem[Casta{\~n}o et~al\mbox{.}(2016)]%
        {castano2016talebox}
\bibfield{author}{\bibinfo{person}{Olatz Casta{\~n}o}, \bibinfo{person}{Ben Kybartas}, {and} \bibinfo{person}{Rafael Bidarra}.} \bibinfo{year}{2016}\natexlab{}.
\newblock \showarticletitle{TaleBox--a mobile game for mixed-initiative story creation}.
\newblock \bibinfo{journal}{\emph{Proceedings of DiGRA-FDG}} (\bibinfo{year}{2016}).
\newblock


\bibitem[Chung et~al\mbox{.}(2022)]%
        {cowritingideagenerating3}
\bibfield{author}{\bibinfo{person}{John Joon~Young Chung}, \bibinfo{person}{Wooseok Kim}, \bibinfo{person}{Kang~Min Yoo}, \bibinfo{person}{Hwaran Lee}, \bibinfo{person}{Eytan Adar}, {and} \bibinfo{person}{Minsuk Chang}.} \bibinfo{year}{2022}\natexlab{}.
\newblock \showarticletitle{TaleBrush: Sketching Stories with Generative Pretrained Language Models}. In \bibinfo{booktitle}{\emph{Proceedings of the 2022 CHI Conference on Human Factors in Computing Systems}} (New Orleans, LA, USA) \emph{(\bibinfo{series}{CHI '22})}. \bibinfo{publisher}{Association for Computing Machinery}, \bibinfo{address}{New York, NY, USA}, Article \bibinfo{articleno}{209}, \bibinfo{numpages}{19}~pages.
\newblock
\showISBNx{9781450391573}
\urldef\tempurl%
\url{https://doi.org/10.1145/3491102.3501819}
\showDOI{\tempurl}


\bibitem[Clark et~al\mbox{.}(2021)]%
        {distinguishablegpt3}
\bibfield{author}{\bibinfo{person}{Elizabeth Clark}, \bibinfo{person}{Tal August}, \bibinfo{person}{Sofia Serrano}, \bibinfo{person}{Nikita Haduong}, \bibinfo{person}{Suchin Gururangan}, {and} \bibinfo{person}{Noah~A. Smith}.} \bibinfo{year}{2021}\natexlab{}.
\newblock \showarticletitle{All That{'}s {`}Human{'} Is Not Gold: Evaluating Human Evaluation of Generated Text}. In \bibinfo{booktitle}{\emph{Proceedings of the 59th Annual Meeting of the Association for Computational Linguistics and the 11th International Joint Conference on Natural Language Processing (Volume 1: Long Papers)}}. \bibinfo{publisher}{Association for Computational Linguistics}, \bibinfo{address}{Online}, \bibinfo{pages}{7282--7296}.
\newblock
\urldef\tempurl%
\url{https://doi.org/10.18653/v1/2021.acl-long.565}
\showDOI{\tempurl}


\bibitem[Compton and Mateas(2015)]%
        {casualcreators}
\bibfield{author}{\bibinfo{person}{Kate Compton} {and} \bibinfo{person}{Michael Mateas}.} \bibinfo{year}{2015}\natexlab{}.
\newblock \showarticletitle{Casual Creators.}. In \bibinfo{booktitle}{\emph{ICCC}}. \bibinfo{pages}{228--235}.
\newblock


\bibitem[Coulianos(2009)]%
        {pacing}
\bibfield{author}{\bibinfo{person}{Filip Coulianos}.} \bibinfo{year}{2009}\natexlab{}.
\newblock \showarticletitle{A Method for Pacing Analysis}.
\newblock \bibinfo{journal}{\emph{Game Career Guide, Oct}}  \bibinfo{volume}{6} (\bibinfo{year}{2009}).
\newblock


\bibitem[Csikszentmihalyi(2000)]%
        {flowtheory}
\bibfield{author}{\bibinfo{person}{Mihaly Csikszentmihalyi}.} \bibinfo{year}{2000}\natexlab{}.
\newblock \bibinfo{booktitle}{\emph{Beyond boredom and anxiety.}}
\newblock \bibinfo{publisher}{Jossey-bass}.
\newblock


\bibitem[Deterding et~al\mbox{.}(2017)]%
        {mixedinitiativeinterface}
\bibfield{author}{\bibinfo{person}{Sebastian Deterding}, \bibinfo{person}{Jonathan Hook}, \bibinfo{person}{Rebecca Fiebrink}, \bibinfo{person}{Marco Gillies}, \bibinfo{person}{Jeremy Gow}, \bibinfo{person}{Memo Akten}, \bibinfo{person}{Gillian Smith}, \bibinfo{person}{Antonios Liapis}, {and} \bibinfo{person}{Kate Compton}.} \bibinfo{year}{2017}\natexlab{}.
\newblock \showarticletitle{Mixed-initiative creative interfaces}. In \bibinfo{booktitle}{\emph{Proceedings of the 2017 CHI Conference Extended Abstracts on Human Factors in Computing Systems}}. \bibinfo{pages}{628--635}.
\newblock


\bibitem[Frayling(1993)]%
        {rtd}
\bibfield{author}{\bibinfo{person}{Christopher Frayling}.} \bibinfo{year}{1993}\natexlab{}.
\newblock \showarticletitle{Research in art and design}.
\newblock \bibinfo{journal}{\emph{Royal College of Art research papers}}  \bibinfo{volume}{1} (\bibinfo{year}{1993}), \bibinfo{pages}{1--5}.
\newblock


\bibitem[Gallotta et~al\mbox{.}(2024)]%
        {llm4game}
\bibfield{author}{\bibinfo{person}{Roberto Gallotta}, \bibinfo{person}{Graham Todd}, \bibinfo{person}{Marvin Zammit}, \bibinfo{person}{Sam Earle}, \bibinfo{person}{Antonios Liapis}, \bibinfo{person}{Julian Togelius}, {and} \bibinfo{person}{Georgios~N Yannakakis}.} \bibinfo{year}{2024}\natexlab{}.
\newblock \showarticletitle{Large Language Models and Games: A Survey and Roadmap}.
\newblock \bibinfo{journal}{\emph{arXiv preprint arXiv:2402.18659}} (\bibinfo{year}{2024}).
\newblock


\bibitem[Halabi et~al\mbox{.}(2019)]%
        {halabi2019assessing}
\bibfield{author}{\bibinfo{person}{Nour Halabi}, \bibinfo{person}{G{\"u}nter Wallner}, {and} \bibinfo{person}{Pejman Mirza-Babaei}.} \bibinfo{year}{2019}\natexlab{}.
\newblock \showarticletitle{Assessing the impact of visual design on the interpretation of aggregated playtesting data visualization}. In \bibinfo{booktitle}{\emph{Proceedings of the Annual Symposium on Computer-Human Interaction in Play}}. \bibinfo{pages}{639--650}.
\newblock


\bibitem[Khandkar(2009)]%
        {opencoding}
\bibfield{author}{\bibinfo{person}{Shahedul~Huq Khandkar}.} \bibinfo{year}{2009}\natexlab{}.
\newblock \showarticletitle{Open coding}.
\newblock \bibinfo{journal}{\emph{University of Calgary}} \bibinfo{volume}{23}, \bibinfo{number}{2009} (\bibinfo{year}{2009}).
\newblock


\bibitem[Kreminski et~al\mbox{.}(2020a)]%
        {wawlt2}
\bibfield{author}{\bibinfo{person}{Max Kreminski}, \bibinfo{person}{Melanie Dickinson}, \bibinfo{person}{Michael Mateas}, {and} \bibinfo{person}{Noah Wardrip-Fruin}.} \bibinfo{year}{2020}\natexlab{a}.
\newblock \showarticletitle{Why Are We Like This?: Exploring writing mechanics for an AI-augmented storytelling game}.
\newblock  (\bibinfo{year}{2020}).
\newblock


\bibitem[Kreminski et~al\mbox{.}(2020b)]%
        {mixedinitiativegame-wawlt}
\bibfield{author}{\bibinfo{person}{Max Kreminski}, \bibinfo{person}{Melanie Dickinson}, \bibinfo{person}{Michael Mateas}, {and} \bibinfo{person}{Noah Wardrip-Fruin}.} \bibinfo{year}{2020}\natexlab{b}.
\newblock \showarticletitle{Why Are We Like This?: The AI Architecture of a Co-Creative Storytelling Game}. In \bibinfo{booktitle}{\emph{Proceedings of the 15th International Conference on the Foundations of Digital Games}} (Bugibba, Malta) \emph{(\bibinfo{series}{FDG '20})}. \bibinfo{publisher}{Association for Computing Machinery}, \bibinfo{address}{New York, NY, USA}, Article \bibinfo{articleno}{13}, \bibinfo{numpages}{4}~pages.
\newblock
\showISBNx{9781450388078}
\urldef\tempurl%
\url{https://doi.org/10.1145/3402942.3402953}
\showDOI{\tempurl}


\bibitem[Kreminski et~al\mbox{.}(2019)]%
        {kreminski2019felt}
\bibfield{author}{\bibinfo{person}{Max Kreminski}, \bibinfo{person}{Melanie Dickinson}, {and} \bibinfo{person}{Noah Wardrip-Fruin}.} \bibinfo{year}{2019}\natexlab{}.
\newblock \showarticletitle{Felt: a simple story sifter}. In \bibinfo{booktitle}{\emph{Interactive Storytelling: 12th International Conference on Interactive Digital Storytelling, ICIDS 2019, Little Cottonwood Canyon, UT, USA, November 19--22, 2019, Proceedings 12}}. Springer, \bibinfo{pages}{267--281}.
\newblock


\bibitem[Kreminski et~al\mbox{.}(2022)]%
        {mixedinitiativegame-looseend}
\bibfield{author}{\bibinfo{person}{Max Kreminski}, \bibinfo{person}{Melanie Dickinson}, \bibinfo{person}{Noah Wardrip-Fruin}, {and} \bibinfo{person}{Michael Mateas}.} \bibinfo{year}{2022}\natexlab{}.
\newblock \showarticletitle{Loose Ends: A Mixed-Initiative Creative Interface for Playful Storytelling}.
\newblock \bibinfo{journal}{\emph{Proceedings of the AAAI Conference on Artificial Intelligence and Interactive Digital Entertainment}} \bibinfo{volume}{18}, \bibinfo{number}{1} (\bibinfo{date}{Oct.} \bibinfo{year}{2022}), \bibinfo{pages}{120--128}.
\newblock
\urldef\tempurl%
\url{https://doi.org/10.1609/aiide.v18i1.21955}
\showDOI{\tempurl}


\bibitem[Kybartas and Bidarra(2015)]%
        {kybartas2015semantic}
\bibfield{author}{\bibinfo{person}{Ben Kybartas} {and} \bibinfo{person}{Rafael Bidarra}.} \bibinfo{year}{2015}\natexlab{}.
\newblock \showarticletitle{A semantic foundation for mixed-initiative computational storytelling}. In \bibinfo{booktitle}{\emph{Interactive Storytelling: 8th International Conference on Interactive Digital Storytelling, ICIDS 2015, Copenhagen, Denmark, November 30-December 4, 2015, Proceedings 8}}. Springer, \bibinfo{pages}{162--169}.
\newblock


\bibitem[Köbis and Mossink(2021)]%
        {distinguishablegpt2}
\bibfield{author}{\bibinfo{person}{Nils Köbis} {and} \bibinfo{person}{Luca~D. Mossink}.} \bibinfo{year}{2021}\natexlab{}.
\newblock \showarticletitle{Artificial intelligence versus Maya Angelou: Experimental evidence that people cannot differentiate AI-generated from human-written poetry}.
\newblock \bibinfo{journal}{\emph{Computers in Human Behavior}}  \bibinfo{volume}{114} (\bibinfo{year}{2021}), \bibinfo{pages}{106553}.
\newblock
\showISSN{0747-5632}
\urldef\tempurl%
\url{https://doi.org/10.1016/j.chb.2020.106553}
\showDOI{\tempurl}


\bibitem[Laclaustra et~al\mbox{.}(2014)]%
        {cowritingfillgap2}
\bibfield{author}{\bibinfo{person}{Iv{\'a}n~M Laclaustra}, \bibinfo{person}{Jos{\'e} Ledesma}, \bibinfo{person}{Gonzalo M{\'e}ndez}, {and} \bibinfo{person}{Pablo Gerv{\'a}s}.} \bibinfo{year}{2014}\natexlab{}.
\newblock \showarticletitle{Kill the Dragon and Rescue the Princess: Designing a Plan-based Multi-agent Story Generator.}. In \bibinfo{booktitle}{\emph{ICCC}}. \bibinfo{pages}{347--350}.
\newblock


\bibitem[Latitude(2023)]%
        {AIDungeon}
\bibfield{author}{\bibinfo{person}{Latitude}.} \bibinfo{year}{2023}\natexlab{}.
\newblock \bibinfo{booktitle}{\emph{AI Dungeon}}.
\newblock
\urldef\tempurl%
\url{https://aidungeon.com/}
\showURL{%
Retrieved November 5, 2023 from \tempurl}


\bibitem[Love(2019)]%
        {Yujianlove}
\bibfield{author}{\bibinfo{person}{Yu~Jian Love}.} \bibinfo{year}{2019}\natexlab{}.
\newblock \bibinfo{booktitle}{\emph{NetEase}}.
\newblock
\urldef\tempurl%
\url{https://yujian.163.com/#/more}
\showURL{%
Retrieved November 4, 2023 from \tempurl}


\bibitem[McCabe and Peterson(1991)]%
        {narrativestructure}
\bibfield{author}{\bibinfo{person}{Allyssa McCabe} {and} \bibinfo{person}{Carole Peterson}.} \bibinfo{year}{1991}\natexlab{}.
\newblock \bibinfo{booktitle}{\emph{Developing narrative structure}}.
\newblock \bibinfo{publisher}{Psychology Press}.
\newblock


\bibitem[Miller(1995)]%
        {miller1995wordnet}
\bibfield{author}{\bibinfo{person}{George~A Miller}.} \bibinfo{year}{1995}\natexlab{}.
\newblock \showarticletitle{WordNet: a lexical database for English}.
\newblock \bibinfo{journal}{\emph{Commun. ACM}} \bibinfo{volume}{38}, \bibinfo{number}{11} (\bibinfo{year}{1995}), \bibinfo{pages}{39--41}.
\newblock


\bibitem[OpenAI(2023)]%
        {GPTDocument}
\bibfield{author}{\bibinfo{person}{OpenAI}.} \bibinfo{year}{2023}\natexlab{}.
\newblock \bibinfo{booktitle}{\emph{OpenAI API Documentation}}.
\newblock
\urldef\tempurl%
\url{https://platform.openai.com/docs/models/overview}
\showURL{%
Retrieved November 4, 2023 from \tempurl}


\bibitem[Osborn et~al\mbox{.}(2015)]%
        {osborn2015playspecs}
\bibfield{author}{\bibinfo{person}{Joseph Osborn}, \bibinfo{person}{Ben Samuel}, \bibinfo{person}{Michael Mateas}, {and} \bibinfo{person}{Noah Wardrip-Fruin}.} \bibinfo{year}{2015}\natexlab{}.
\newblock \showarticletitle{Playspecs: Regular expressions for game play traces}. In \bibinfo{booktitle}{\emph{Proceedings of the AAAI conference on artificial intelligence and interactive digital entertainment}}, Vol.~\bibinfo{volume}{11}. \bibinfo{pages}{170--176}.
\newblock


\bibitem[Radford et~al\mbox{.}(2019)]%
        {gpt2}
\bibfield{author}{\bibinfo{person}{Alec Radford}, \bibinfo{person}{Jeffrey Wu}, \bibinfo{person}{Rewon Child}, \bibinfo{person}{David Luan}, \bibinfo{person}{Dario Amodei}, \bibinfo{person}{Ilya Sutskever}, {et~al\mbox{.}}} \bibinfo{year}{2019}\natexlab{}.
\newblock \showarticletitle{Language models are unsupervised multitask learners}.
\newblock \bibinfo{journal}{\emph{OpenAI blog}} \bibinfo{volume}{1}, \bibinfo{number}{8} (\bibinfo{year}{2019}), \bibinfo{pages}{9}.
\newblock


\bibitem[Reed et~al\mbox{.}(2014)]%
        {reed2014ice}
\bibfield{author}{\bibinfo{person}{Aaron~A Reed}, \bibinfo{person}{Jacob Garbe}, \bibinfo{person}{Noah Wardrip-Fruin}, {and} \bibinfo{person}{Michael Mateas}.} \bibinfo{year}{2014}\natexlab{}.
\newblock \showarticletitle{Ice-Bound: Combining richly-realized story with expressive gameplay.}. In \bibinfo{booktitle}{\emph{FDG}}.
\newblock


\bibitem[Roemmele and Gordon(2015)]%
        {cowritingideageneration1}
\bibfield{author}{\bibinfo{person}{Melissa Roemmele} {and} \bibinfo{person}{Andrew~S. Gordon}.} \bibinfo{year}{2015}\natexlab{}.
\newblock \showarticletitle{Creative Help: A Story Writing Assistant}. In \bibinfo{booktitle}{\emph{Interactive Storytelling}}, \bibfield{editor}{\bibinfo{person}{Henrik Schoenau-Fog}, \bibinfo{person}{Luis~Emilio Bruni}, \bibinfo{person}{Sandy Louchart}, {and} \bibinfo{person}{Sarune Baceviciute}} (Eds.). \bibinfo{publisher}{Springer International Publishing}, \bibinfo{address}{Cham}, \bibinfo{pages}{81--92}.
\newblock
\showISBNx{978-3-319-27036-4}


\bibitem[Samuel et~al\mbox{.}(2016)]%
        {mixedinitativegames-buddy}
\bibfield{author}{\bibinfo{person}{Ben Samuel}, \bibinfo{person}{Michael Mateas}, {and} \bibinfo{person}{Noah Wardrip-Fruin}.} \bibinfo{year}{2016}\natexlab{}.
\newblock \showarticletitle{The Design of Writing Buddy: A Mixed-Initiative Approach Towards Computational Story Collaboration}. In \bibinfo{booktitle}{\emph{Interactive Storytelling}}, \bibfield{editor}{\bibinfo{person}{Frank Nack} {and} \bibinfo{person}{Andrew~S. Gordon}} (Eds.). \bibinfo{publisher}{Springer International Publishing}, \bibinfo{address}{Cham}, \bibinfo{pages}{388--396}.
\newblock
\showISBNx{978-3-319-48279-8}


\bibitem[Samuel et~al\mbox{.}(2015)]%
        {samuel2015ensemble}
\bibfield{author}{\bibinfo{person}{Ben Samuel}, \bibinfo{person}{Aaron~A Reed}, \bibinfo{person}{Paul Maddaloni}, \bibinfo{person}{Michael Mateas}, {and} \bibinfo{person}{Noah Wardrip-Fruin}.} \bibinfo{year}{2015}\natexlab{}.
\newblock \showarticletitle{The ensemble engine: Next-generation social physics}. In \bibinfo{booktitle}{\emph{Proceedings of the Tenth International Conference on the Foundations of Digital Games (FDG 2015)}}. \bibinfo{pages}{22--25}.
\newblock


\bibitem[Schell(2008)]%
        {schell2008artgamedesign}
\bibfield{author}{\bibinfo{person}{Jesse Schell}.} \bibinfo{year}{2008}\natexlab{}.
\newblock \bibinfo{booktitle}{\emph{The Art of Game Design: A book of lenses}}.
\newblock \bibinfo{publisher}{CRC press}.
\newblock


\bibitem[Sharma et~al\mbox{.}(2019)]%
        {SnakeGame}
\bibfield{author}{\bibinfo{person}{Shubham Sharma}, \bibinfo{person}{Saurabh Mishra}, \bibinfo{person}{Nachiket Deodhar}, \bibinfo{person}{Akshay Katageri}, {and} \bibinfo{person}{Parth Sagar}.} \bibinfo{year}{2019}\natexlab{}.
\newblock \showarticletitle{Solving The Classic Snake Game Using AI}. In \bibinfo{booktitle}{\emph{2019 IEEE Pune Section International Conference (PuneCon)}}. \bibinfo{pages}{1--4}.
\newblock
\urldef\tempurl%
\url{https://doi.org/10.1109/PuneCon46936.2019.9105796}
\showDOI{\tempurl}


\bibitem[Shi et~al\mbox{.}(2022)]%
        {cowriteedit1}
\bibfield{author}{\bibinfo{person}{Shuming Shi}, \bibinfo{person}{Enbo Zhao}, \bibinfo{person}{Duyu Tang}, \bibinfo{person}{Yan Wang}, \bibinfo{person}{Piji Li}, \bibinfo{person}{Wei Bi}, \bibinfo{person}{Haiyun Jiang}, \bibinfo{person}{Guoping Huang}, \bibinfo{person}{Leyang Cui}, \bibinfo{person}{Xinting Huang}, \bibinfo{person}{Cong Zhou}, \bibinfo{person}{Yong Dai}, {and} \bibinfo{person}{Dongyang Ma}.} \bibinfo{year}{2022}\natexlab{}.
\newblock \bibinfo{title}{Effidit: Your AI Writing Assistant}.
\newblock
\newblock
\urldef\tempurl%
\url{https://doi.org/10.48550/ARXIV.2208.01815}
\showDOI{\tempurl}


\bibitem[Sun et~al\mbox{.}(2023)]%
        {1001night}
\bibfield{author}{\bibinfo{person}{Yuqian Sun}, \bibinfo{person}{Zhouyi Li}, \bibinfo{person}{Ke Fang}, \bibinfo{person}{Chang~Hee Lee}, {and} \bibinfo{person}{Ali Asadipour}.} \bibinfo{year}{2023}\natexlab{}.
\newblock \showarticletitle{Language as reality: a co-creative storytelling game experience in 1001 nights using generative AI}. In \bibinfo{booktitle}{\emph{Proceedings of the AAAI Conference on Artificial Intelligence and Interactive Digital Entertainment}}, Vol.~\bibinfo{volume}{19}. \bibinfo{pages}{425--434}.
\newblock


\bibitem[Sun et~al\mbox{.}(2022)]%
        {sun2022bringing}
\bibfield{author}{\bibinfo{person}{Yuqian Sun}, \bibinfo{person}{Xuran Ni}, \bibinfo{person}{Haozhen Feng}, \bibinfo{person}{Ray LC}, \bibinfo{person}{Chang~Hee Lee}, {and} \bibinfo{person}{Ali Asadipour}.} \bibinfo{year}{2022}\natexlab{}.
\newblock \showarticletitle{Bringing stories to life in 1001 nights: A co-creative text adventure game using a story generation model}. In \bibinfo{booktitle}{\emph{International Conference on Interactive Digital Storytelling}}. Springer, \bibinfo{pages}{651--672}.
\newblock


\bibitem[Swanson and Gordon(2012)]%
        {cowritingideageneration2}
\bibfield{author}{\bibinfo{person}{Reid Swanson} {and} \bibinfo{person}{Andrew~S. Gordon}.} \bibinfo{year}{2012}\natexlab{}.
\newblock \showarticletitle{Say Anything: Using Textual Case-Based Reasoning to Enable Open-Domain Interactive Storytelling}.
\newblock \bibinfo{journal}{\emph{ACM Trans. Interact. Intell. Syst.}} \bibinfo{volume}{2}, \bibinfo{number}{3}, Article \bibinfo{articleno}{16} (\bibinfo{date}{sep} \bibinfo{year}{2012}), \bibinfo{numpages}{35}~pages.
\newblock
\showISSN{2160-6455}
\urldef\tempurl%
\url{https://doi.org/10.1145/2362394.2362398}
\showDOI{\tempurl}


\bibitem[Tocci(2008)]%
        {conflict}
\bibfield{author}{\bibinfo{person}{Jason Tocci}.} \bibinfo{year}{2008}\natexlab{}.
\newblock \showarticletitle{“You Are Dead. Continue?”: Conflicts and Complements in Game Rules and Fiction}.
\newblock \bibinfo{journal}{\emph{Eludamos: Journal for Computer Game Culture}} \bibinfo{volume}{2}, \bibinfo{number}{2} (\bibinfo{year}{2008}), \bibinfo{pages}{187--201}.
\newblock


\bibitem[Wiethof et~al\mbox{.}(2021)]%
        {wiethof2021designinggamifiedmixed}
\bibfield{author}{\bibinfo{person}{Christina Wiethof}, \bibinfo{person}{Navid Tavanapour}, {and} \bibinfo{person}{Eva Bittner}.} \bibinfo{year}{2021}\natexlab{}.
\newblock \showarticletitle{Designing and evaluating a collaborative writing process with gamification elements: Toward a framework for gamifying collaboration processes}.
\newblock \bibinfo{journal}{\emph{AIS Transactions on Human-Computer Interaction}} \bibinfo{volume}{13}, \bibinfo{number}{1} (\bibinfo{year}{2021}), \bibinfo{pages}{38--61}.
\newblock


\bibitem[Wu et~al\mbox{.}(2022)]%
        {pixelization}
\bibfield{author}{\bibinfo{person}{Zongwei Wu}, \bibinfo{person}{Liangyu Chai}, \bibinfo{person}{Nanxuan Zhao}, \bibinfo{person}{Bailin Deng}, \bibinfo{person}{Yongtuo Liu}, \bibinfo{person}{Qiang Wen}, \bibinfo{person}{Junle Wang}, {and} \bibinfo{person}{Shengfeng He}.} \bibinfo{year}{2022}\natexlab{}.
\newblock \showarticletitle{Make Your Own Sprites: Aliasing-Aware and Cell-Controllable Pixelization}.
\newblock \bibinfo{journal}{\emph{ACM Trans. Graph.}} \bibinfo{volume}{41}, \bibinfo{number}{6}, Article \bibinfo{articleno}{193} (\bibinfo{date}{nov} \bibinfo{year}{2022}), \bibinfo{numpages}{16}~pages.
\newblock
\showISSN{0730-0301}
\urldef\tempurl%
\url{https://doi.org/10.1145/3550454.3555482}
\showDOI{\tempurl}


\bibitem[Xi et~al\mbox{.}(2021)]%
        {xi-etal-2021-kuileixi}
\bibfield{author}{\bibinfo{person}{Yadong Xi}, \bibinfo{person}{Xiaoxi Mao}, \bibinfo{person}{Le Li}, \bibinfo{person}{Lei Lin}, \bibinfo{person}{Yanjiang Chen}, \bibinfo{person}{Shuhan Yang}, \bibinfo{person}{Xuhan Chen}, \bibinfo{person}{Kailun Tao}, \bibinfo{person}{Zhi Li}, \bibinfo{person}{Gongzheng Li}, \bibinfo{person}{Lin Jiang}, \bibinfo{person}{Siyan Liu}, \bibinfo{person}{Zeng Zhao}, \bibinfo{person}{Minlie Huang}, \bibinfo{person}{Changjie Fan}, {and} \bibinfo{person}{Zhipeng Hu}.} \bibinfo{year}{2021}\natexlab{}.
\newblock \showarticletitle{{K}ui{L}ei{X}i: a {C}hinese Open-Ended Text Adventure Game}. In \bibinfo{booktitle}{\emph{Proceedings of the 59th Annual Meeting of the Association for Computational Linguistics and the 11th International Joint Conference on Natural Language Processing: System Demonstrations}}, \bibfield{editor}{\bibinfo{person}{Heng Ji}, \bibinfo{person}{Jong~C. Park}, {and} \bibinfo{person}{Rui Xia}} (Eds.). \bibinfo{publisher}{Association for Computational Linguistics}, \bibinfo{address}{Online}, \bibinfo{pages}{175--184}.
\newblock
\urldef\tempurl%
\url{https://doi.org/10.18653/v1/2021.acl-demo.21}
\showDOI{\tempurl}


\bibitem[Yang et~al\mbox{.}(2022)]%
        {cowriting-aiasactivewriter}
\bibfield{author}{\bibinfo{person}{Daijin Yang}, \bibinfo{person}{Yanpeng Zhou}, \bibinfo{person}{Zhiyuan Zhang}, \bibinfo{person}{Toby Jia-Jun Li}, {and} \bibinfo{person}{Ray LC}.} \bibinfo{year}{2022}\natexlab{}.
\newblock \showarticletitle{AI as an Active Writer: Interaction strategies with generated text in human-AI collaborative fiction writing}. In \bibinfo{booktitle}{\emph{Joint Proceedings of the ACM IUI Workshops}}, Vol.~\bibinfo{volume}{10}.
\newblock


\bibitem[Zhang et~al\mbox{.}(2023)]%
        {Zhang_2023_ICCV_Controlnet}
\bibfield{author}{\bibinfo{person}{Lvmin Zhang}, \bibinfo{person}{Anyi Rao}, {and} \bibinfo{person}{Maneesh Agrawala}.} \bibinfo{year}{2023}\natexlab{}.
\newblock \showarticletitle{Adding Conditional Control to Text-to-Image Diffusion Models}. In \bibinfo{booktitle}{\emph{Proceedings of the IEEE/CVF International Conference on Computer Vision (ICCV)}}. \bibinfo{pages}{3836--3847}.
\newblock


\bibitem[Zhao(0)]%
        {cowriteedit2}
\bibfield{author}{\bibinfo{person}{Xin Zhao}.} \bibinfo{year}{0}\natexlab{}.
\newblock \showarticletitle{Leveraging Artificial Intelligence (AI) Technology for English Writing: Introducing Wordtune as a Digital Writing Assistant for EFL Writers}.
\newblock \bibinfo{journal}{\emph{RELC Journal}} \bibinfo{volume}{0}, \bibinfo{number}{0} (\bibinfo{year}{0}), \bibinfo{pages}{00336882221094089}.
\newblock
\urldef\tempurl%
\url{https://doi.org/10.1177/00336882221094089}
\showDOI{\tempurl}
\showeprint{https://doi.org/10.1177/00336882221094089}


\end{thebibliography}

\end{document}